\documentclass[preprint,showpacs,preprintnumbers,amsmath,amssymb]{revtex4}

\usepackage{graphicx}% Include figure files
\usepackage{dcolumn}% Align table columns on decimal point
\usepackage{bm}% bold math
\def\ä{\"{a}}
\def\ü{\"{u}}
\def\ö{\"{o}}
\def\Ä{\"{A}}
\def\Ü{\"{U}}
\def\Ö{\"{O}}

\begin{document}

\preprint{APS/123-QED}

\title{Simultaneous current-, force- and work function measurement with atomic resolution}% Force line breaks with \\

\author{M. Herz}
\author{Ch. Schiller}
\author{F. J. Giessibl}
\email{Franz.Giessibl@physik.uni-augsburg.de}
\author{J. Mannhart}

%\altaffiliation[Also at ]{}%Lines break automatically or can be forced with \\
%\author{}%

\affiliation{%
Universit\"at Augsburg, Institute of Physics, Electronic
Correlations and Magnetism, Experimentalphysik VI,
Universit\"atsstrasse 1, D-86135 Augsburg,
Germany.\homepage{http://www.Physik.Uni-Augsburg.DE/exp6}
}%

\date{Submitted to Applied Physics Letters Sep 22 2004, revised version Jan 19 2005}% It is always \today, today,

\begin{abstract}
The local work function of a surface determines the spatial decay
of the charge density at the Fermi level normal to the surface.
Here, we present a method that enables simultaneous measurements
of local work function and tip-sample forces. A combined dynamic
scanning tunneling microscope and atomic force microscope is used
to measure the tunneling current between an oscillating tip and
the sample in real time as a function of the cantilever's
deflection. Atomically resolved work function measurements on a
silicon (111)-($7\times 7$) surface are presented and related to
concurrently recorded tunneling current- and force- measurements.
\end{abstract}

\maketitle

When two metallic electrodes are separated by a small vacuum gap
and a bias voltage $V_t$ is applied, a quantum mechanical
tunneling current flows \cite{Teague:1978,Binnig:1982a,Chen:1993}.
For metallic electrodes, the tunneling current increases
approximately by a factor of 10 for each distance reduction of
100\,pm. This sharp distance dependence is key to the atomic
resolution capability of the scanning tunneling microscope (STM)
\cite{Binnig:1982}. In one dimension, the decay of the tunneling
current with distance is roughly given by
\begin{equation}\label{Iz}
    I = I_0 \exp{(-2{\kappa}z)}
\end{equation}
with
\begin{equation}\label{Phi}
    \kappa = \sqrt{2m\Phi/}\hbar
\end{equation}
where $m$ is the mass of an electron and $\Phi$ is the local work
function or barrier height \cite{Chen:1993}. As Binnig has shown
early in the development of STM, this exponential dependence holds
over at least 0.5\,nm or so for tunneling currents ranging from a
few hundred pA to a few $\mu$A \cite{Binnig:1982a}. Due to the
atomic structure of matter, $\kappa$ is also a function of the
lateral positions $x$ and $y$ as well as $z$ (see Fig. 1) for very
small tunneling distances.

A straightforward method for measuring $\kappa(x,y,z)$ is to stop
the lateral scan over a specific atom position and perform a
$I(z)$ measurement. However, this method is time consuming and
prone to errors from creep and drift of the piezoelectric scanner.
Pethica et al. \cite{Pethica:1993} have extended \lq
current-imaging-tunneling-spectroscopy\rq (CITS) by Hamers et al.
\cite{Hamers:1986}, an ac-method for current vs. voltage
spectroscopy where a small ac-voltage is added to $V_t$ to recover
the density of states. In Pethica et al.'s  method, the tip is
oscillated at a fixed frequency $f$ (typically on the order of a
few kHz) according to $\vec{x}=\vec{x_0}+\vec{A}\cos(2\pi f t)$
\cite{Pethica:1993}. For this situation, Eq. \ref{Iz} needs to be
generalized to
\begin{equation}\label{Ix}
    I = I_0 \exp{(-2{\vec{\kappa}}\vec{x})}.
\end{equation}
The natural logarithm of the normalized current is then given by
\begin{equation}\label{lnIx}
    \ln(I/I_0) = -2{\vec{\kappa}}\vec{x_0}-2\vec{\kappa}\vec{A}\cos(2\pi f t).
\end{equation}
In general, we define an apparent decay \lq constant\rq
$\tilde{\kappa}=\eta/(2A)$, where $\eta$ is the AC-component of
$\ln(I/I_0)$. For the case described by \mbox{Eq. \ref{lnIx}},
$\eta=2\vec{\kappa}\vec{A}$.

Atomic force microscopy (AFM) has progressed rapidly in the past
years \cite{Garcia:2002,Giessibl:2003}, and a combination with
other techniques like STM \cite{Herz:2003a} or Kelvin Probe
Mircoscopy \cite{Kitamura:1998} became feasible. Here, we combine
AFM with Pethica's method by using a qPlus sensor
\cite{Giessibl:1998} where the STM tip is mounted on a vibrating
cantilever. The oscillation frequency is no longer fixed in this
case, but varies by a frequency shift $\Delta f$ as determined by
the atomic forces acting between tip and sample. The frequency
shift data can then be related to forces \cite{Giessibl:2003}, and
simultaneous measurements of forces and decay constants are
possible.

Because the atomically resolved STM image is influenced by the
atomic and subatomic structure of the tip and sample wavefunctions
\cite{Chen:1993,Herz:2003a,Herz:2003b}, these structures also
influence the decay constant images \cite{Pethica:1993}. The
tunneling current can be calculated with a plane-wave expansion of
the surface wavefunctions \cite{Chen:1993}, ultimately depending
on the superposition of atomic orbitals
\cite{Chen:1993,Herz:2003a,Herz:2003b}.

\begin{figure}
\includegraphics[width=7cm]{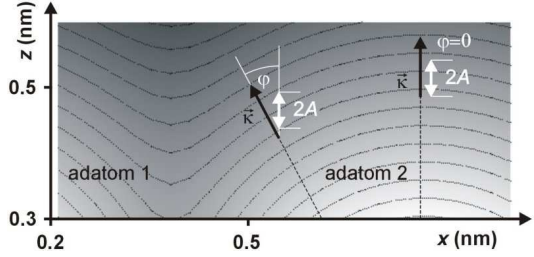}
\caption{\label{Fig1} Schematic view of $\ln{(I/I_0)}$ above two
silicon adatoms located at $z=0$ and $z=0.77$\,nm of a silicon
(111)-($7\times 7$) surface. The surface is oriented perpendicular
to the $z$-axis. The current distribution between each of the
silicon states and a s-tip is approximately given by
$I\propto(z/r)^2 \exp{(-2\vec{\kappa}\vec{r})}$ \cite{Chen:1993}.
The amplitude of the vertical oscillation is $A$. $\varphi$ is the
angle between the oscillation direction and $\vec{\kappa}$. For
decreasing $\cos(\varphi)$, the decay strength of the current in
vertical direction also decreases (see text). Note that generally
$\vec{\kappa}$ is \textit{not} parallel to
$\vec{\nabla}\ln{(I/I_0)}$ because of the angular factor of the
wave function. The distance of the contour lines corresponds to a
current increase of 1.82.}
\end{figure}

For the atomic basis functions at the Fermi level, the exponential
radial factor is $\exp{(-\vec{\kappa}\vec{r})}$, where
$\vec{\kappa}=\kappa \vec{r}/r$ and $\vec{r}$ is the position
vector with respect to the nucleus. When the tip moves by
$\vec{\alpha}$, the corresponding conductivity distribution
changes according to $\exp{(-2\vec{\kappa}\vec{\alpha})}$
\cite{Chen:1993}. Fig. \ref{Fig1} shows a cross-section of a model
distribution of $\ln{(I/I_0)}$ for two atomic silicon states of a
silicon (111)-($7\times 7$) surface, with the nuclei on the
\mbox{$x$-axis} at $z=0$, separated by \mbox{0.77\,nm}. The
current dependence $I\propto(z/r)^2 \exp{(-2\vec{\kappa}\vec{r})}$
is a good fit for tunneling between a tip (s-tip) and each of the
sp$^3$-adatom states \cite{Chen:1993}. The tip is assumed to
oscillate vertically. $\varphi$ is the angle between the
oscillation direction and $\vec{\kappa}$. For
$\cos(\varphi)\rightarrow 1$, $\eta$ approaches
$2\vec{\kappa}\vec{A}=2\kappa A$ in this model. Therefore, an
\textit{apparent} barrier height
$\Phi_{\mathrm{app}}=\hbar^2\tilde{\kappa}^2/(2m)$ can be
extracted from the value of $\tilde{\kappa}$ at the position of
the single atom. For decreasing $\cos(\varphi)$, $\tilde{\kappa}$
also decreases in the theoretical model, since $\vec{A}$ is no
longer parallel to the direction of the fastest decay of the
current. In the model, odd higher powers of
$[\ln(I/I_0)](q)=[\ln(I/I_0)](\vec{x_0}+q\vec{A}/A)$ influence the
value of $\tilde{\kappa}$ as well.

\begin{figure}
\includegraphics[width=7cm]{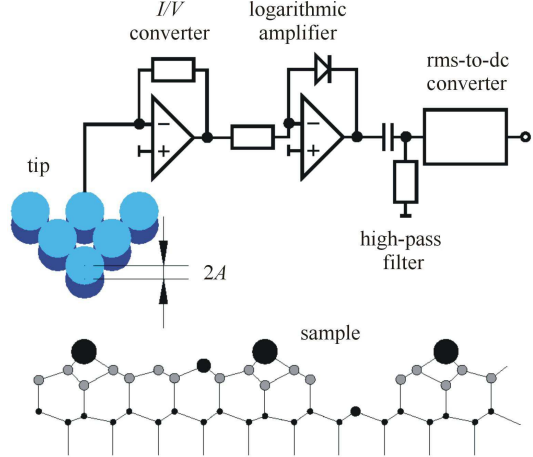}
\caption{\label{Fig2} {Principle of dynamic workfunction
measurement. The tip is mounted on a quartz cantilever (not shown
here, see \cite{Giessibl:1998} for details) with a stiffness of
1800 N/m which oscillates at a fixed amplitude $A \approx 0.1$ nm.
The unperturbed resonance frequency is $f_0 \approx 20$\,kHz. Due
to the sinusoidal oscillation of the cantilever, the tunneling
current is strongly modulated at a frequency $f \approx f_0$ (see
Fig. 3).}}
\end{figure}

In this publication, we present dynamic STM/AFM measurements
\cite{Herz:2003a} performed in ultrahigh vacuum at a pressure of
$p\approx10^{-8}$\,Pa and ambient temperature $T\approx$ 300\,K. A
silicon (111)-($7\times 7$) surface is probed by a tungsten tip
that is mounted on a qPlus force sensor \cite{Giessibl:1998}. The
qPlus sensor used here has a stiffness of $k=1800$\,N/m and an
eigenfrequency of 15487\,Hz. The oscillation amplitude of the
cantilever needs to be approximately below \mbox{0.1 nm} such that
$I$ is above the noise level of the current amplifier even when
the tip is at the far point of the surface. Operation at such
small amplitudes poses a challenge as amplitude fluctuations tend
to increase with decreasing amplitude, and only the use of
cantilevers with a very large stiffness decreases amplitude
fluctuations to an acceptable level \cite{Giessibl:2003}. The
experimental setup is shown in \mbox{Fig. \ref{Fig2}}.

\begin{figure}
\includegraphics[width=7cm]{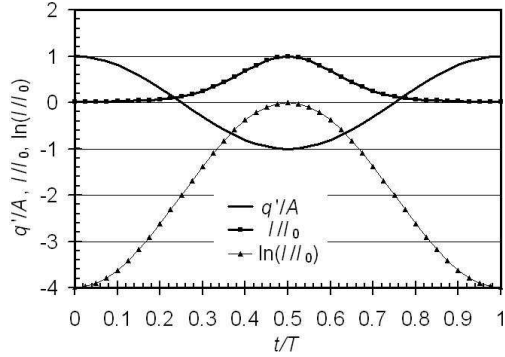}
\caption{\label{Fig3} {Tip deflection divided by amplitude,
current divided by maximum current and natural logarithm of
$I/I_0$ as a function of time $t$ for $t=0$ to $t=1/f$. If the
oscillation amplitude is large compared to the decay length of the
tunneling current, the current appears as a series of Gaussian
peaks.}}
\end{figure}

Fig. \ref{Fig3} shows a simulation of the excursion of the tip
mounted on a qPlus cantilever, the normalized tunneling current
$I$ and the natural logarithm of $I$ for an ideal exponential
current-distance dependence.

\begin{figure}
\includegraphics[width=7cm]{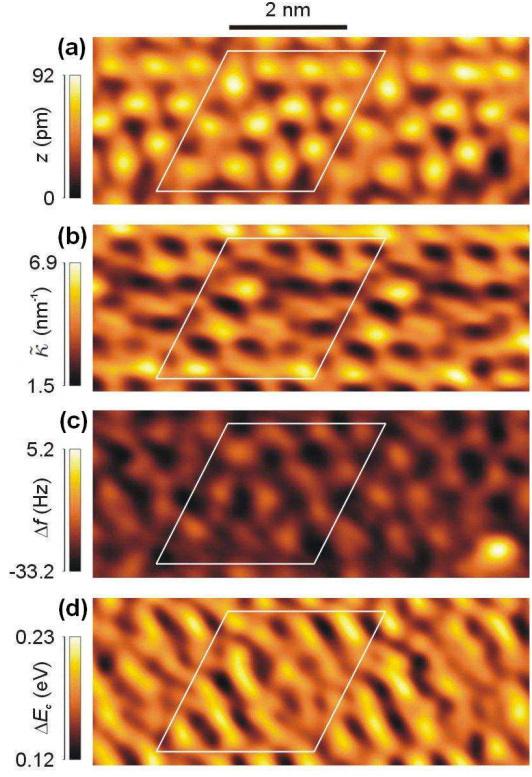}
\caption{\label{Fig4} Simultaneous measurement of topography
(constant average tunneling current) ({\bf a}), $\tilde{\kappa}$
({\bf b}), frequency shift $\Delta f$ ({\bf c}) and damping signal
$\Delta E_c$ per oscillation cycle ({\bf d}). The image was
acquired in the dynamic STM mode with a tungsten tip on the
silicon (111)-($7\times 7$) surface and a qPlus sensor
\cite{Giessibl:1998} with a quality factor of \mbox{$Q= 2240$}.
The surface unit cell is indicated with white diamonds. Sample
bias: \mbox{2.0 V}, average tunneling current \mbox{$I_{av} =$ 0.8
nA}, amplitude \mbox{$A =$ 0.1 nm}.}
\end{figure}

Fig. \ref{Fig4} shows a simultaneous measurement of topography
(constant average current), $\tilde{\kappa}$, frequency shift
$\Delta f$ and damping signal $\Delta E_c$ per oscillation cycle.
The angle between $\vec{A}$ and the surface normal was
$\theta\approx 20^\circ$. The cantilever-amplitude was
$A=100$\,pm$\pm 4$\,pm. At a single adatom defect site in the
right bottom of Fig. \ref{Fig4} (c), the frequency shift is
positive because of a very small tip-sample distance, where
repulsive forces are acting
\cite{Herz:2003a,Herz:2003b,Hembacher:2003}. The presence of a
single atomic defect proves that the tip has a single front atom
and multitip effects can be ruled out. The average damping signal
is \mbox{$\overline{\Delta E_c}=$\,170 meV} and exceeds the
internal damping of the cantilever by \mbox{$\approx$ 10\,meV}. On
an atomic scale, $\tilde{\kappa}$ shows a strong variation. The
maximum value of $\tilde{\kappa}$ is \mbox{$\approx$ 65\%} of the
nominal value of $\kappa$ for this tip-sample system
\mbox{(workfunction $\Phi\approx$ 4.5\,eV, \cite{Ashcroft:1981})}.
The average of $\tilde{\kappa}$ is less than 50\% of the nominal
value of $\kappa$. This lowering of the measured average
$\tilde{\kappa}$ can not be explained with the hypothetic geometry
factor $\cos(\theta)$ for the case of tunneling between parallel
surfaces, which would yield only 6\% of the deviation. However, a
considerable lowering and the strong spatial variation are in
agreement with earlier measurements using a tungsten tip on a
silicon (113) surface \cite{Pethica:1993}. A lowering was also
observed in the first current-distance measurements at a voltage
of \mbox{60 mV} \cite{Binnig:1982a}. The variation of
$\tilde{\kappa}$ at a subatomic scale can be explained with
geometry effects in the orbital model for the tunneling process
\cite{Chen:1993,Herz:2003a,Herz:2003b} as shown above. At some
adatoms of the silicon (111)-($7\times 7$) unit-cell,
$\tilde{\kappa}$ shows a maximum as expected in the simple orbital
model. The lateral displacement of the maxima of $\tilde{\kappa}$
with respect to the topographic maxima is in very good agreement
with the results of a model calculation for a s-tip on the silicon
surface at the experimental conditions (\mbox{$\theta\approx
20^\circ$}). However, along the surface projection of $\vec{A}$
(vertical direction in the paper plane of Fig. \ref{Fig4}),
$\tilde{\kappa}$ decreases faster than in the model, probably
caused by displacements of tip- and sample atoms at the small
tip-sample distance. The variations in the maximal values of
$\tilde{\kappa}$ between the four types of adatoms on the silicon
(111)-($7\times 7$) surface point to variations in the local work
function.

Partially, the lowering of the maximum value of $\tilde{\kappa}$
may be caused by the spatial extension of the tip atom or atom
cluster. An extended tip results in a blurring of the image,
compared to the theoretical images and should result in a lower
value of $\tilde{\kappa}$. In proximity to the sample the
polynomial radial factor of the atomic basis functions can
additionally cause a lowering of $\tilde{\kappa}$ compared to
$\kappa$. Finally, the finite voltage causes changes in the
barrier shape between tip and sample, which might also lower the
value of $\tilde{\kappa}$.

The observed features may be a consequence of the structure of the
surface wavefunctions which are probed in dynamic STM
\cite{Chen:1993,Herz:2003a,Herz:2003b}. However, displacements of
tip and sample atoms and the formation of chemical bonds also have
to be considered because of the very small tip-sample-distance
\cite{Hofer:2001}. Figure 4 (c) shows that the frequency shift is
less negative on top of the Si adatoms than between the adatoms.
Because adatoms exert strong attractive forces on the tip before
contact \cite{Giessibl:1995}, we conclude that the short-range
forces acting between tip and sample in Fig. 4 are already
repulsive. Bond formation can additionally cause variations in the
decay constant images on an atomic scale. By which extent the
measured variation is caused by mechanical deformations of the tip
and/or sample, the formation of a chemical bond or by geometry
effects solely because of the structure of the surface electron
states can not be determined at this stage.

In conclusion, we have presented simultaneous measurements of the
tunneling current, the work function and the tip-sample forces
with atomic resolution. Future improvements of the technique could
be achieved by utilizing force sensors with even higher spring
constants, such that smaller oscillation amplitudes become
possible. Orienting the sensor oscillation exactly perpendicular
to the surface would facilitate the interpretation of the data. We
suggest that differences in the local barrier heights might help
to identify the charge state or species of single atoms. The
topographic corrugation can be increased dramatically using higher
angular momentum tip states
\cite{Chen:1993,Herz:2003a,Herz:2003b}. For atomic scale
measurements with such tips, an enhanced corrugation of the work
function images and higher harmonics of $\ln(I/I_0)$ is expected.

\begin{acknowledgments}
We thank S. Hembacher for discussions. This work is supported by
the Bundesministerium f\"{u}r Bildung und Forschung (project
EKM13N6918).
\end{acknowledgments}

\newpage

\bibliography{2004fjg}

\begin{thebibliography}{16}
\expandafter\ifx\csname natexlab\endcsname\relax\def\natexlab#1{#1}\fi
\expandafter\ifx\csname bibnamefont\endcsname\relax
  \def\bibnamefont#1{#1}\fi
\expandafter\ifx\csname bibfnamefont\endcsname\relax
  \def\bibfnamefont#1{#1}\fi
\expandafter\ifx\csname citenamefont\endcsname\relax
  \def\citenamefont#1{#1}\fi
\expandafter\ifx\csname url\endcsname\relax
  \def\url#1{\texttt{#1}}\fi
\expandafter\ifx\csname urlprefix\endcsname\relax\def\urlprefix{URL }\fi
\providecommand{\bibinfo}[2]{#2}
\providecommand{\eprint}[2][]{\url{#2}}

\bibitem[{\citenamefont{Binnig et~al.}(1982{\natexlab{a}})\citenamefont{Binnig,
  Rohrer, Gerber, and Weibel}}]{Binnig:1982a}
\bibinfo{author}{\bibfnamefont{G.}~\bibnamefont{Binnig}},
  \bibinfo{author}{\bibfnamefont{H.}~\bibnamefont{Rohrer}},
  \bibinfo{author}{\bibfnamefont{C.}~\bibnamefont{Gerber}}, \bibnamefont{and}
  \bibinfo{author}{\bibfnamefont{E.}~\bibnamefont{Weibel}},
  \bibinfo{journal}{Appl. Phys. Lett.} \textbf{\bibinfo{volume}{40}},
  \bibinfo{pages}{178} (\bibinfo{year}{1982}{\natexlab{a}}).

\bibitem[{\citenamefont{Chen}(1993)}]{Chen:1993}
\bibinfo{author}{\bibfnamefont{C.~J.} \bibnamefont{Chen}},
  \emph{\bibinfo{title}{Introduction to Scanning Tunneling Microscopy}}
  (\bibinfo{publisher}{{}Oxford University Press, New York},
  \bibinfo{year}{1993}).

\bibitem[{\citenamefont{Teague}(1986)}]{Teague:1978}
\bibinfo{author}{\bibfnamefont{E.~C.} \bibnamefont{Teague}},
  \bibinfo{journal}{Thesis at North Texas University (1978), reprinted in J.
  Research of the National Bureau of Standards} \textbf{\bibinfo{volume}{91}},
  \bibinfo{pages}{171} (\bibinfo{year}{1986}).

\bibitem[{\citenamefont{Binnig et~al.}(1982{\natexlab{b}})\citenamefont{Binnig,
  Rohrer, Gerber, and Weibel}}]{Binnig:1982}
\bibinfo{author}{\bibfnamefont{G.}~\bibnamefont{Binnig}},
  \bibinfo{author}{\bibfnamefont{H.}~\bibnamefont{Rohrer}},
  \bibinfo{author}{\bibfnamefont{C.}~\bibnamefont{Gerber}}, \bibnamefont{and}
  \bibinfo{author}{\bibfnamefont{E.}~\bibnamefont{Weibel}},
  \bibinfo{journal}{Phys. Rev. Lett.} \textbf{\bibinfo{volume}{49}},
  \bibinfo{pages}{57} (\bibinfo{year}{1982}{\natexlab{b}}).

\bibitem[{\citenamefont{Pethica et~al.}(1993)\citenamefont{Pethica, Knall, and
  Wilson}}]{Pethica:1993}
\bibinfo{author}{\bibfnamefont{J.~B.} \bibnamefont{Pethica}},
  \bibinfo{author}{\bibfnamefont{J.}~\bibnamefont{Knall}}, \bibnamefont{and}
  \bibinfo{author}{\bibfnamefont{J.~H.} \bibnamefont{Wilson}},
  \bibinfo{journal}{Institute of Physics Conf. Ser., London}
  \textbf{\bibinfo{volume}{134}}, \bibinfo{pages}{597} (\bibinfo{year}{1993}).

\bibitem[{\citenamefont{Hamers et~al.}(1986)\citenamefont{Hamers, Tromp, and
  Demuth}}]{Hamers:1986}
\bibinfo{author}{\bibfnamefont{R.~J.} \bibnamefont{Hamers}},
  \bibinfo{author}{\bibfnamefont{R.~M.} \bibnamefont{Tromp}}, \bibnamefont{and}
  \bibinfo{author}{\bibfnamefont{J.~E.} \bibnamefont{Demuth}},
  \bibinfo{journal}{Phys. Rev. Lett.} \textbf{\bibinfo{volume}{56}},
  \bibinfo{pages}{1972–1975} (\bibinfo{year}{1986}).

\bibitem[{\citenamefont{Garcia and Perez}(2002)}]{Garcia:2002}
\bibinfo{author}{\bibfnamefont{R.}~\bibnamefont{Garcia}} \bibnamefont{and}
  \bibinfo{author}{\bibfnamefont{R.}~\bibnamefont{Perez}},
  \bibinfo{journal}{Surf. Sci. Rep.} \textbf{\bibinfo{volume}{47}},
  \bibinfo{pages}{197} (\bibinfo{year}{2002}).

\bibitem[{\citenamefont{Giessibl}(2003)}]{Giessibl:2003}
\bibinfo{author}{\bibfnamefont{F.~J.} \bibnamefont{Giessibl}},
  \bibinfo{journal}{Rev. Mod. Phys.} \textbf{\bibinfo{volume}{75}},
  \bibinfo{pages}{949} (\bibinfo{year}{2003}).

\bibitem[{\citenamefont{Herz et~al.}(2003)\citenamefont{Herz, Giessibl, and
  Mannhart}}]{Herz:2003a}
\bibinfo{author}{\bibfnamefont{M.}~\bibnamefont{Herz}},
  \bibinfo{author}{\bibfnamefont{F.~J.} \bibnamefont{Giessibl}},
  \bibnamefont{and} \bibinfo{author}{\bibfnamefont{J.}~\bibnamefont{Mannhart}},
  \bibinfo{journal}{{}Phys. Rev. B} \textbf{\bibinfo{volume}{68}},
  \bibinfo{pages}{045301} (\bibinfo{year}{2003}).

\bibitem[{\citenamefont{Kitamura and Iwatsuki}(1998)}]{Kitamura:1998}
\bibinfo{author}{\bibfnamefont{S.}~\bibnamefont{Kitamura}} \bibnamefont{and}
  \bibinfo{author}{\bibfnamefont{M.}~\bibnamefont{Iwatsuki}},
  \bibinfo{journal}{{}Appl. Phys. Lett.} \textbf{\bibinfo{volume}{72}},
  \bibinfo{pages}{3154} (\bibinfo{year}{1998}).

\bibitem[{\citenamefont{Giessibl}(1998)}]{Giessibl:1998}
\bibinfo{author}{\bibfnamefont{F.~J.} \bibnamefont{Giessibl}},
  \bibinfo{journal}{{}Appl. Phys. Lett.} \textbf{\bibinfo{volume}{73}},
  \bibinfo{pages}{3956} (\bibinfo{year}{1998}).

\bibitem[{\citenamefont{Herz}(2003)}]{Herz:2003b}
\bibinfo{author}{\bibfnamefont{M.}~\bibnamefont{Herz}}, Ph.D. thesis,
  \bibinfo{school}{University of Augsburg, Germany} (\bibinfo{year}{2003}).

\bibitem[{\citenamefont{Hembacher et~al.}(2003)\citenamefont{Hembacher,
  Giessibl, Mannhart, and Quate}}]{Hembacher:2003}
\bibinfo{author}{\bibfnamefont{S.}~\bibnamefont{Hembacher}},
  \bibinfo{author}{\bibfnamefont{F.~J.} \bibnamefont{Giessibl}},
  \bibinfo{author}{\bibfnamefont{J.}~\bibnamefont{Mannhart}}, \bibnamefont{and}
  \bibinfo{author}{\bibfnamefont{C.~F.} \bibnamefont{Quate}},
  \bibinfo{journal}{{}Proc. Natl. Acad. Sci. (USA)}
  \textbf{\bibinfo{volume}{100}}, \bibinfo{pages}{12539}
  (\bibinfo{year}{2003}).

\bibitem[{\citenamefont{Ashcroft and Mermin}(1981)}]{Ashcroft:1981}
\bibinfo{author}{\bibfnamefont{N.~W.} \bibnamefont{Ashcroft}} \bibnamefont{and}
  \bibinfo{author}{\bibfnamefont{N.~D.} \bibnamefont{Mermin}},
  \emph{\bibinfo{title}{Solid State Physics}} (\bibinfo{publisher}{Saunders
  College, Philadelphia}, \bibinfo{year}{1981}).

\bibitem[{\citenamefont{Hofer et~al.}(2001)\citenamefont{Hofer, Fisher, Wolkow,
  and Grutter}}]{Hofer:2001}
\bibinfo{author}{\bibfnamefont{W.~A.} \bibnamefont{Hofer}},
  \bibinfo{author}{\bibfnamefont{A.~J.} \bibnamefont{Fisher}},
  \bibinfo{author}{\bibfnamefont{R.~A.} \bibnamefont{Wolkow}},
  \bibnamefont{and} \bibinfo{author}{\bibfnamefont{P.}~\bibnamefont{Grutter}},
  \bibinfo{journal}{Phys. Rev. Lett.} \textbf{\bibinfo{volume}{87}},
  \bibinfo{pages}{236104} (\bibinfo{year}{2001}).

\bibitem[{\citenamefont{Giessibl}(1995)}]{Giessibl:1995}
\bibinfo{author}{\bibfnamefont{F.~J.} \bibnamefont{Giessibl}},
  \bibinfo{journal}{Science} \textbf{\bibinfo{volume}{267}},
  \bibinfo{pages}{68} (\bibinfo{year}{1995}).

\end{thebibliography}

\end{document}